\documentclass[aps,epsfig,float,twocolumn,prl]{revtex4}
\newcommand{\be}{\begin{equation}}
\newcommand{\ee}{\end{equation}}

\usepackage{graphicx}
\usepackage{amsmath}
\usepackage{amstext}
\usepackage{amssymb}
\usepackage{amsfonts}
\usepackage{amsbsy}

\begin{document}
\title{Community Detection as an Inference Problem}
\author{M. B. Hastings}
\affiliation{Center for Nonlinear Studies and Theoretical Division,
Los Alamos National
Laboratory, Los Alamos, NM 87545
}
\date{April 6, 2006}
\begin{abstract}
We express community detection as an inference problem of determining
the most likely arrangement of communities.  We then apply
belief propagation and mean-field theory to this problem, and show that
this leads to fast, accurate algorithms for community detection.
\vskip2mm
\end{abstract}
\maketitle
%]
Community detection is a well-studied problem in networks\cite{nr}.
This is the problem of dividing a network into communities,
such that nodes within the same community tend to be connected by links,
while those within different communities tend not to be connected by
links.
This problem has applications in understanding the structure of social and
biological networks\cite{applic}, while the closely related graph partitioning problem
discussed below has applications in parallel processing, to allocate
assignments to different processors while minimizing interprocessor
communication.

Unfortunately, despite all this interest, there is no formal definition of the
problem.  Instead, each author tends to define communities as being whatever
is found by a particular community detection algorithm\cite{cdr2}.  In this work,
we exploit a standard method of testing communities, the four groups
test\cite{ng}, to express community detection as an inference, or maximum
likelihood problem.  This leads to a derivation of a Potts model similar to
those derived previously on phenomenological grounds\cite{potts}.

To solve this inference problem, we must find the ground state of the
Potts model.  To do this, we turn to the techniques of {\it belief propagation}\cite{belief},
also known as sum-product, and mean-field theory.
Belief propagation was originally developed to perform decoding in a certain
class of error correcting codes, called low-density parity check codes\cite{bpb}.  In this
problem, one has a sender and receiver communicating over a
noisy channel, and the receiver must determine which of all the possible messages is the most
likely.  This problem can be mapped onto finding the ground state of a spin system on
a particular graph\cite{sourlas}.  The belief propagation algorithm exploits the fact that the graph
has a low density of loops to solve this problem, in a manner similar to the
famous Bethe-Peierls\cite{bethe} solution for the thermodynamics of a spin system on
a Bethe lattice.

We will find that the resulting belief propagation algorithm for community detection is
highly accurate on the four groups problem, while also performing well
on other test networks.  We then discuss the scaling of computational time with system
size, extensions of this algorithm, and other problems.

{\it Inference Problem---}
To express community detection as an inference problem
we consider the following method, often used to test community detection 
algorithms\cite{ng}.
We invent a network as follows: we consider $N$ nodes, divided
into $q$ different communities with $n_k$, $k=1,...,q$ nodes in each community.
We consider each pair of nodes in turn.  We connect those
nodes with probability $p_{in}$ if they lie within the same community and
$p_{out}$ if they lie within different communities.  We then run the community
detection algorithm and see if it correctly assigns nodes to communities.
Let $q_i$ be the initial assignment of a community to node $i$.
The probability that a given graph arises from this procedure is equal to
\begin{eqnarray}
\label{pqi}
p(\{q_i\})=
\\ \nonumber
\left( \prod_{k=1}^q (1-p_{in})^{n_k^2/2-n_k/2} \right)
\left( \prod_{k=1}^{q-1} \prod_{l=k+1}^q 
(1-p_{out})^{n_k n_l}
\right)
\times
\\ \nonumber
\left( \prod_{<ij>} (\frac{p_{in}}{1-p_{in}})^{\delta_{q_i,q_j}} \right)
\left( \prod_{<ij>} (\frac{p_{out}}{1-p_{out}})^{1-\delta_{q_i,q_j}} \right)
\end{eqnarray}
where $\prod_{<ij>}$ denotes a product over pairs of vertices $i,j$ connected by
an edge.
To verify the correctness of this formula, first consider the
case in which there are no edges at all in the graph.  Then,
the probability is given correctly by the first two products
of Eq.~(\ref{pqi}).  Adding edges to the graph changes the probability
as given by the second two products in Eq.~(\ref{pqi}).

We now consider a given graph and formulate the problem as follows:
find the most likely community assignment.
Following Bayes' theorem, given a graph, the probability that any given
community assignment is the ``correct" assignment is proportional to the probability
$p(\{q_i\})$ multiplied by the a priori probability of having a given $n_k$.
Throughout this paper we assume the a priori probability of a given $n_k$ is constant,
and thus the optimal community assignment maximizes the probability in Eq.~(\ref{pqi}); any
non-constant priori probability can be easily incorporated by adding additional terms.

We rewrite Eq.~(\ref{pqi}) as an exponential:
\be
\label{pqie}
p(\{q_i\})=
c \exp[\sum_{<ij>} J \delta_{q_i,q_j}]
\exp[\sum_{i\neq j} J' \delta_{q_i,q_j}/2],
\ee
where
$\sum_{<ij>}$ denotes a sum over pairs of $i,j$ connected by an edge in the
graph, and
$c=\exp[\log(1-p_{out})N(N-1)/2]
\exp[\sum_{<ij>} \log(p_{out}/(1-p_{out})]$, and with
\begin{eqnarray}
\label{constants}
J=\log[(p_{in}(1-p_{out}))/((1-p_{in})p_{out})],\\ \nonumber
J'=\log[(1-p_{in})/(1-p_{out})].
\end{eqnarray}
The factor of $1/2$ in Eq.~(\ref{pqie}) is to avoid double counting.
Eq.~(\ref{pqie}) presents the probability as a Potts model problem with
combined short- and long-range interactions, with coupling constants
$J,J'$.  The problem of community detection is then reduced to
finding the ground state of this Potts model.  Assuming $p_{in}>p_{out}$,
the short-range interactions are ferromagnetic, favoring the
assignment of neighboring nodes to the same community, while the
long-range interactions are anti-ferromagnetic and prevent one from
simply taking all nodes to lie within the same community.
The problem of finding the ground state is very closely
related to the NP-complete problem of graph partitioning, to break a graph up
into partitions, minimizing the number of edges connecting partitions and
minimizing the difference in number of nodes between partitions.

{\it Belief Propagation---}
Having arrived at Eq.~(\ref{pqie}) we have a very similar problem to that
studied in \cite{potts}.
Instead of using Monte Carlo methods to find the ground state, we adopt the method
of belief propagation, which we believe to be more efficient for many of the community
detection problems that arise.  Indeed, for the inference problems which arise 
many in
error correcting codes, belief propagation is the most efficient method.

We begin by taking a mean-field approximation for the long-range interactions, justified
for large $N$.  We approximate
$p(\{q_i\})\approx
p_{mft}(\{q_i\})$, where
\begin{eqnarray}
p_{mft}(\{q_i\})\equiv
Z^{-1} \exp[\sum_{<ij>} J \delta_{q_i,q_j}] 
\exp[\sum_{i} h_i(q_i)]
\end{eqnarray}
where $Z$ is a normalization, and for $r=1...q$ we define
$h_i(r)=J' N \rho(r)-J' p_i(r)$
 and
$\rho(r)=\sum_i p_i(r)/N$, with 
\be
p_i(r)=\frac{\sum_{\{q_j\}, q_i=r} p_{mft}(\{q_j\})}{\sum_{\{q_j\}} p_{mft}(\{q_j\})},
\ee
so that $p_i(r)$
is the probability in the mean-field approximation that node $i$ belongs to community
$r$.
These are a set of self-consistent
equations for $p_i(q)$.

We will find that solving these equations, at least in the belief propagation
approximation below, leads to a spontaneous symmetry breaking: for sufficiently
large $J,J'$, the probability $p_i(q)$ depends on $q$.  Of course, given any solution
which breaks symmetry, one can arrive at other valid solutions by relabeling the
communities.  We then make a ``nodewise" maximum a posteriori probability (MAP) approximation:
for each node $i$, we compute $p_i(q)$ and then assign the node $i$ to the community
$q$ which maximizes $p_i(q)$.  This is an approximation: the set of community assignments which
maximizes $p(\{q_i\})$ may not be given by maximizing the probability for each node separately.
However, similar approximations work very well for error correcting codes\cite{bpb}, and
these approximations are justified for large $J,J'$.

To compute the $p_i(q)$, we apply belief propagation.  Suppose the graph forms a tree,
with no loops.  Then, the problem of solving for $p_i(q)$ given $h_i(q)$ can be solved:
for each pair of nodes $i,j$ connected by an edge, we define $p_{ij}(r)$ to be probability
$p_i$ for the network modified by removing the edge connecting $i$ to $j$.  That is,
\be
p_{ij}(r)=\frac{\sum_{\{q_k\}, q_i=r} \exp(-J \delta_{q_i,q_j})
p_{mft}(\{q_k\})}{\sum_{\{q_k\}} \exp(-J \delta_{q_i,q_j}) p_{mft}(\{q_k\})},
\ee
as $\exp(-J \delta_{q_i,q_j}) p_{mft}(\{q_k\})$ is the probability distribution on
the network with the edge removed.
Then, for a tree-like structure, the Bethe-Peierls solution gives
\be
\label{bppi}
p_i(r)=Z_i^{-1} \exp[h_i(r)]\prod_{j}^{i,j \, n.n.} 
\bigl[ \exp(J) p_{ji}(r)+(1-p_{ji}(r) \bigr] ,
\ee
where $Z_i$ is chosen so that $\sum_{r=1}^q p_i(r)=1$, and where the product $\prod_{j}^{i,j \, n.n.}$
is taken over all nodes $j$ which are connected to node $i$.  Note that
$\exp(J) p_{ji}(r)+(1-p_{ji}(r))=\sum_s p_{ji}(s) (\delta_{r,s} \exp(J)+(1-\delta_{r,s}))$.
Similarly,
\be
\label{bppij}
p_{ij}(r)=Z_{ij}^{-1} \exp[h_i(r)]\prod_{k\neq j}^{i,k \, n.n.}
\bigl[\exp(J) p_{ki}(r)+(1-p_{ki}(r)\bigr],
\ee
where again we choose $Z_{ij}$ such that $\sum_{r=1}^q p_{ij}(r)=1$.

These belief propagation equations are exact for a tree.  If the network is not a tree, however,
belief propagation can still be very effective, especially if the density of loops
is small.  Of course, not only may the graph have loops, but the long-range 
interactions also induce loops.  However, for large $N$, the mean-field approximation we have
used is justified for the long-range interactions, and as discussed below the
procedure works well even for test networks with loops.

We can simplify the algorithm
by replacing the belief propagation equations by a set of naive mean-field equations:
for each site we track only the beliefs $p_i(q_i)$ and iterate the equations
\be
\label{mfteq}
p_i(r)=Z_i^{-1} \prod_{j,<ij>} \exp[h_i(r)] \bigl(\exp(J) p_j(r)+ (1-p_j(r)) \bigr).
\ee
This method is most appropriate for networks with large $z$ where it
becomes faster than the belief propagation
method by roughly a factor $z$ as there are fewer equations to solve; also, as we will see, for large $z$ the mean-field
approximation used here performs comparably to belief propagation.

There are many ways to solve the belief propagation equations.  We choose to initialize
each of the $p_i(r)=1/q+\epsilon_i(r)$, where $\epsilon_i(r)$ is chosen randomly and is small,
and similarly for each of the $p_{ij}(r)$.  We then perform a fixed number of iterations:
on each iteration we first randomly select one directed edge $i,j$ and then
update the corresponding function
$p_{ij}(r)$ by replacing its current value with $0.75$ times its current value plus
$0.25$ times the value given by solving Eq.~(\ref{bppij}).  We then randomly select one
node $i$ and replace the function $p_i(r)$ by its $0.75$ times its current value plus
$0.25$ times the value given by solving Eq.~(\ref{bppi}).  After updating $p_i(r)$ we
update $\rho(r)$ and $m_i(r)$, thus solving the belief propagation and mean-field equations
simultaneously.  Other relaxation methods may prove better in certain applications.

{\it Applications---}
We have tested this algorithm on several problems.  We first consider the four groups
problem.  In this case, we take a randomly generated
network of $N=128$ nodes, divided into four communities
of 32 nodes each.  
After generating the network, we run the community detection algorithm, to find the
most likely assignment of communities.  We measure the algorithm's performance by
determining the fraction of nodes whose community it identifies correctly.  There is some
arbitrariness in how this fraction of correctly identified nodes is defined, which is
connected with the arbitrariness in the labeling of communities: one can permute the
community labels as desired, given the Potts symmetry of Eq.~(\ref{pqie}).  To
resolve this arbitrariness, we follow the stringent definition adopted in \cite{newmandef}
for the accuracy.

We choose $p_{in}$ and $p_{out}$ so that each node has
on average $z_{in}$ connections to other nodes in the same community and $z_{out}$
connections to nodes in different communities.  We pick $z_{in}+z_{out}=16$, and
consider a range of values of $z_{in}$, the average number of intra-community links.
For large $z_{in}$,
the community detection problem is easier, as the effect of the community
structure is much more clear.  

\begin{figure}[!t]
\begin{center}
\leavevmode
\includegraphics[width=2.5in]{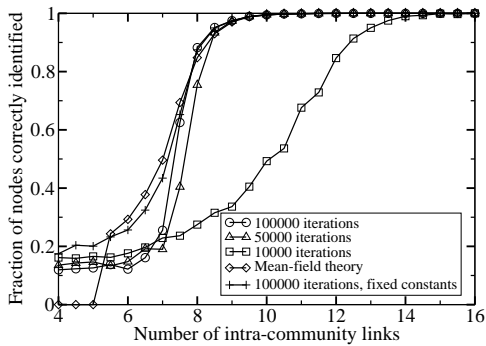}
\caption{Accuracy of belief propagation and mean-field algorithms.  Circles, squares, and triangles
represent different numbers of iterations, using constants from Eq.~(\ref{constants}).
Pluses represent a different, fixed choice of constants described in text.}
\end{center}  
\end{figure} 

The results of this procedure are shown in Fig.~1.  As seen, 
with 100000 iterations, the algorithm
is highly accurate.  With 50000 iterations, a slight decrease in accuracy is
noticed for small
$z_{in}$, and the accuracy goes down significantly at $10000$ iterations (by increasing
to $200000$ iterations, a slight improvement is noted for $z_{in}<8$).  For the first
three curves, we used the choice of constants in Eq.~(\ref{constants}).  Since these constants
depend on the given $z_{in},z_{out}$, something which may not be known for an arbitrary network,
we repeated the algorithm with a particular fixed choice of constants,
$\exp[J]=6, \exp[NJ']=10^{-10}$.  This choice of constants was chosen completely
arbitrarily, but as seen the algorithm still works well, with a much
higher accuracy than the Newman-Girvan algorithm.  However, the Newman-Girvan algorithm
has some advantages in terms of picking the most optimal number of different communities
into which to divide the network, while the present algorithm takes the number of communities
as an input.  We ignore, however, the a priori knowledge that each community has 32
nodes.

Finally, we have tested the mean-field algorithm.  As seen, this is almost as accurate as
the belief propagation algorithm for $z_{in}\geq 8$, and actually performs better
for $z_{in}\leq 7.5$.  We believe that the improved performance is a result of the
fact that the mean-field equations more easily break symmetry.  Tests on networks with
a lower coordination number showed the difference more clearly, as given in Fig.~2 for
a network with four groups and $z_{in}+z_{out}=4$.

\begin{figure}[!t]
\begin{center}
\leavevmode
\includegraphics[width=2.5in]{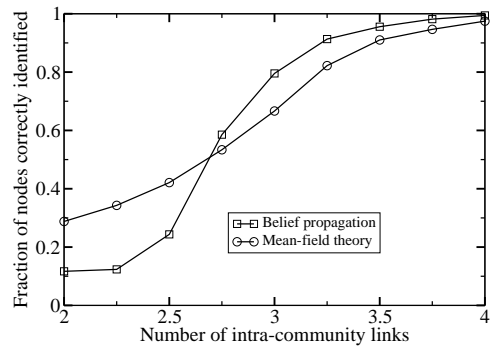}
\caption{Accuracy of belief propagation and mean-field algorithms for $z_{in}+z_{out}=4$.}
\end{center}  
\end{figure} 

We have also tested the belief propagation algorithm on simple networks, such as dividing $N$
nodes arranged on a straight line with connections between nearest
neighbors on the line into two different
communities, as well as the Zachary karate club network\cite{zach}.
The relaxation of the algorithm needed to be done slightly more slowly than
described above (ten randomly chosen nodes were relaxed before each edge
was relaxed), and some care was taken on the constants due to the lower
coordination number in order to obtain convergence:
with poor choices the beliefs $p_i(r)$ oscillated randomly.  After finding these
constants on the straight line network,
the algorithm was tested on the Zachary network, and
identified the communities accurately, with 
one error of placing node $10$ in the wrong community, as labeled
in the figure in \cite{ng}.
However, this node has one link to each of the two
communities and so the network gives nodes
no obvious preference for either community.
Interestingly, after convergence of the equations,
for almost
all nodes $i$, the maximum over $r$ of $p_i(r)$ was
greater than $0.99$, except node $9$ where it was only $0.98$ and
node $10$ where it was only $0.87$; these two nodes have
roughly equal connections to both communities.

{\it Discussion---}
We have expressed community detection as an inference problem, providing a formulation
of the problem in statistical mechanical terms.  We have then applied the belief propagation
method to solve the resulting statistical problem.  The results are accurate, however
there are a number of questions that should be addressed as well as possible extensions.
First, there is some questions about picking the constants $J,J'$.  In some cases, especially
on test networks with a low coordination number, a poor choice of constants leads to either a lack
of convergence of the belief propagation equations, or else convergence to a solution in which
$p_i(r)\rightarrow 1/q$ so that the spontaneous symmetry breaking is absent.
In both cases the algorithm performs poorly at finding the communities.
The former case requires a slower relaxation of the equations, while the latter case requires an increase
in the constants $J,J'$.  Fortunately, both of these cases can be detected by looking at
the $p_i(r)$ as the algorithm runs, and then corrected, so that the algorithm warns
of its possible failure in these cases.  We did not find any case in
which the belief propagation equations converged to a poor solution which spontaneously
broke symmetry.

The next question is the scaling of the algorithm with system size.  Accurate results were
found with $50000$ iterations.  Since each iteration updates one directed edge, there are roughly
$50000/(128*16)\approx 24$ iterations per edge.  Even with $10000$ iterations, or roughly $5$ per
edge, some information is found.  For a general network, we expect that if there are few
links between communities, then the number of iterations required per edge will be proportional
to the phase-ordering time for a given community under the appropriate dynamics.  This phase
ordering time typically scales\cite{pho} as some power of the relevant length scale for a community,
and for many networks this length scale is of order $\log(N)$.  We thus expect for a network
with average coordination number $z$ that the time will typically be of order $zN\log(N)^{\alpha}$
for some $\alpha$.

There are a number of possible ways of modifying the algorithm, also.  It may be desirable to
incorporate additional a priori knowledge about the communities, or to study the modularity of
the different divisions, as in \cite{ng}.
In some cases, the belief
propagation equations may be reduced to linear programming\cite{lp}.  
A final interesting question relates to our need for spontaneous symmetry breaking.
It would be desirable to be able to go beyond belief propagation using methods such
as \cite{loopex}.  However, doing this may cause us to lose the symmetry breaking, and thus
the MAP approximation may need to be replaced.  
Survey propagation\cite{survey} may help in this regard, as it
allows us to consider a distribution of magnetic fields for each site,
each possibility corresponding to a different symmetry breaking.
Even without these extensions, however,
the present algorithm leads to useful results.

{\it Acknowledgements---} I thank M. Chertkov and
E. Ben-Naim for useful discussions.
This work was supported by US DOE W-7405-ENG-36.

\end{document}